\documentclass{aa}
\usepackage{times,psfig}
\usepackage{epsfig}

\def \deg{^\circ}

\def \hcm {\hbox {\ifmmode $ cm$^{-2}\else cm$^{-2}$\fi}}

\def\approxgt{\mathrel{\hbox{\rlap{\lower.55ex \hbox {$\sim$}}
        \kern-.3em \raise.4ex \hbox{$>$}}}}
\def\approxlt{\mathrel{\hbox{\rlap{\lower.55ex \hbox {$\sim$}}
        \kern-.3em \raise.4ex \hbox{$<$}}}}

\begin{document}

\title{Measuring mass and angular momentum of black holes with high-frequency quasi-periodic oscillations}

\author{B. Aschenbach
}
\offprints{Bernd Aschenbach\\ (bra@mpe.mpg.de)}

\institute{Max-Planck-Institut f\"{u}r extraterrestrische Physik,
P.O. Box 1312, Garching bei M\"{u}nchen, 85741, Germany
}
\date{Received ...; Accepted ...  }

\abstract{3:2 and/or 3:1 twin high frequency quasi-periodic oscillations (HFQPOs)
have been measured for the three microquasars
GRO J1655-40, XTE J1550-564 and GRS 1915+105. 
For a test particle orbiting a rotating black on a stable circular orbit there exist two different orbits
at which the vertical and radial epicyclic oscillations are in either a 3:1 or 3:2 parametric resonance for any 
choice of the black hole angular momentum $a$. If the two orbits are required to be frequency commensurable Keplerian orbits
there is only one solution for the two orbit radii and $a$. This model predicts that the  microquasars have the same $a$, and it 
predicts their black hole masses on the basis of the measured HFQPOs in agreement with the dynamically 
determined masses. Application of this model to the Galactic Center black hole Sgr\,A* using the recently measured 
QPOs 
(Genzel et al., 2003, Aschenbach et al., 2004) 
leads to a black hole mass of $(3.28 \pm 0.13)\times$10$\sp 6 M\sb{\odot}$, and the same $a$  as for the 
microquasars.   
The possibility that all four sources have $a=0.99616$ suggests that this value is the 
upper limit of $a$ imposed by general relativity. 
The same value for the lower orbit radius and the same value for $a$ are also suggested by an analysis 
of the general relativistic expression for the radial gradient of the orbital velocity, 
 which changes sign in a narrow annular region around the lower orbit when $a>0.9953$.

\keywords{accretion discs, QPOs, relativity,  
 -- Galaxy: center 
 -- Galaxy: nucleus
 -- Galaxy: fundamental parameters
 -- X-rays: individuals: Sgr\,A* 
 -- X-rays: individuals: GRO J1655-40
 -- X-rays: individuals: XTE J1550-564
 -- X-rays: individuals: GRS 1915+105
 -- X-rays: individuals: XTE J1859-226
 -- X-rays: general}
}
\titlerunning{A model for QPO's of disk-accreting black holes }
\authorrunning{Aschenbach }
\maketitle 

\section{Introduction}
High-frequency qasi-periodic oscillations (HFQPO) have been observed in quite a number of X-ray binaries containing a 
compact object, i.e. neutron stars or black holes. HFQPOs are considered to originate in the accretion disk around the 
compact object. A special group is obviously set up by 
the galactic microquasars GRO J1655-40, XTE J1550-564 and GRS 1915+105. They are 
special because they do not  show just one frequency but twin frequencies at a fixed ratio of 3:2 
(Abramowicz \& 
Klu\'zniak, 2001, Remillard et al., 2002), which are 
450, 300 Hz (Strohmayer 2001a, Remillard et al., 1999), 
276, 184 Hz (Remillard et al., 2002) and 
168, 113 Hz (Remillard et al., 2003, McClintock \& Remillard, 2004), respectively.  
Out of those three objects   
XTE J1550-564 might have an additional power spectral density peak at 92 Hz, the significance of which is modest, however (Remillard et al., 2002).

Around the time the discoveries were made, it was suggested independently that some 
resonance effect in the accretion disk could give rise 
to such a 3:2 frequency behaviour 
(Klu\'zniak \& Abramowicz, 2001a, b). 
 There are three fundamental oscillation modes for a test particle orbiting a black hole, 
which produce the orbital or azimuthal (Kepler) frequency, the radial (in the orbital plane) epicyclic frequency and the vertical 
or polar (perpendicular to the orbital plane) epicyclic frequency. Initially it has been suggested that the 3:2 frequency 
ratio is due to a coupling of orbital and radial frequency (Abramowicz \& 
Klu\'zniak, 2001, 
Schnittman \& Bertschinger, 2004). However, the proponents of the resonance theory then concentrated on a 
parametric 3:2 resonance between the polar and the radial epicyclic modes (Klu\'zniak \& 
Abramowicz 2002, Abramowicz et al. 2002, 2003, Abramowicz \& Klu\'zniak 2003, 
Klu\'zniak \&
Abramowicz 2003, Rebusco, 2004), and Remillard et al. (2002) found solutions for a 3:2 
frequency ratio for each of the three possible 
couplings, i.e. orbital/radial, orbital/polar and radial/polar.     
The method of associating a frequency to one specific oscillation mode has in principle the benefit that from frequency 
measurements the mass ${M}$, the angular momentum ${a}$ (the Kerr parameter) of the black hole and the radial 
position of the orbiting test particle in the accretion disk can be determined. But with the measurement of  
even several frequencies the relevant equations remain underdetermined and more information is needed. So, in general 
the mass, which has been determined from the dynamics of the binary orbit for each of the three microquasars, has been made 
use of and a value of ${a}$ could be determined, which, however, is different for each of the three coupling possibilities. 
Furthermore, as Remillard et al. (2002) have pointed out on their findings on XTE J1550-564 that the frequency ratio might not be just 
3:2 but that 3:1 is another possibility, one ends up with a choice for ${a}$ of one out of six, still adopting 
${M}$ from dynamical measurements. 
There may be a possibility to resolve this ambiguity by introducing another 
constraint in the resonance model. 

I have investigated whether both a 3:2 and a 3:1 parametric resonance between the vertical and radial epicyclic frequencies 
would exist, 
of course at two different orbits, but that these two orbits are commensurable orbits. The commensurability is meant in the 
traditional sense that the Kepler frequencies of the two orbits come in ratios of natural numbers. Actually, there exists 
such a configuration and there is only one solution, so that the Kepler frequency of the inner orbit is three times the 
Kepler frequency of the outer orbit. Furthermore, this solution allows only one value for the angular momentum, which is 
${a~=~0.99616}$. The mass of the black hole ${M}$ is uniquely determined after a choice for one of the two possible 
orbits has been made. It is likely that the discriminator between the orbits is the mass accretion rate. 
In any case the two possible values for ${M}$ differ by a factor of 1.5 only.    
The masses of the three black holes in GRO J1655-40, XTE J1550-564 and GRS 1915+105, predicted by this model on the 
basis of their measured HFQPOs, agree in each case with the dynamically determined masses within their measurement 
uncertainty range. I have applied this model also to the Galactic Center black hole, Sgr\,A*,
making use of the one quasi-period recently published by Genzel et al. (2003) and the quasi-periods published by 
Aschenbach et al. (2004). Again, the predicted black hole mass agrees very well with the dynamically determined 
mass (section 3). The possible relevance of the 3:2 resonance model to the periodicities reported for Sgr\,A* has been mentioned 
in the papers of Klu\'zniak, Abramowicz \& Lee (2003) and Abramowicz et al. (2004).   

I have checked the general relativistic expressions for the motion of a test particle orbiting a rotating black hole using  
the canonical Boyer-Lindquist functions for some unexpected behaviour near the two orbits and in the 
angular momentum range around ${a~=~0.99616}$. Energy and angular momentum of the test particle remain 
unconspicuous, but the orbital velocity does not. Whereas for any ${a~<~0.9953}$ the orbital velocity 
is monotonically increasing with decreasing radius over the full range of the radius down to the marginally 
stable orbit, for ${a~>~0.9953}$ the orbital velocity starts to decrease with decreasing radius over a small 
radial range before it again rises with decreasing radius. This annular range of decreasing orbital 
velocity happens to be located where the 3:1 parametric resonance of the vertical and radial epicyclic 
frequencies appears for ${a~=~0.99616}$. To me it seems that this finding in not a chance coincidence but is backing 
the resonance model above. The detailed physics still have to be explored. Further details are given in 
section 4. Some discussion about the implications is presented in section 5. 

\section{The model}

There are three fundamental cyclic gravitational modes associated with black hole accretion disks 
(Nowak \& Lehr \cite{NL1998}, Merloni et al. \cite{MVSB1999}),
which are the Kepler frequency ($\Omega\sb{\rm K}$),
the disk perturbation frequencies in vertical and radial
direction called vertical ($\Omega\sb{\rm V}$) and
radial ($\Omega\sb{\rm R}$) epicyclic frequency. Each  frequency
depends on the central, black hole mass $M$,
its angular momentum $a$ and the radial distance $r$ from the center.
Equations \ref{eq:1} to \ref{eq:3} show the relations, for which the
standard notation of c=G=1 is used. Physical length scales
 are in units of GM/c$\sp 2$ and angular frequencies
$\Omega$ are in
units of c$\sp 3$/GM. $r$ = 1 is defined as the gravitational radius r$\sb{\rm g}$.
The conversion from observed linear frequency $\nu$ measured in Hz to the normalized $\Omega$ is obtained through 
$\nu = \Omega\times$c$\sp 3$/(2~$\pi$~GM) with c, G, M in cgs units. 
 
\begin{equation}\label{eq:1}
$$\Omega\sb{\rm K} = (r\sp{3/2} +  a)\sp{-1}$$
\end{equation}
\begin{equation}\label{eq:2}
$$\Omega\sb{\rm V}\sp 2 = \Omega\sb{\rm K}\sp 2 ~ (1 - {{4 a}\over{ r\sp{3/2}}} + {3 a\sp2\over{ r\sp 2}}) $$
\end{equation}
\begin{equation}\label{eq:3}
$$\Omega\sb{\rm R}\sp 2 = \Omega\sb{\rm K}\sp 2 ~ (1 - {6\over{r}} + {{8 a}\over{r\sp{3/2}}} - {3 a\sp2\over{ r\sp 2}}) $$
\end{equation}

\begin{figure}[t]
\psfig{figure=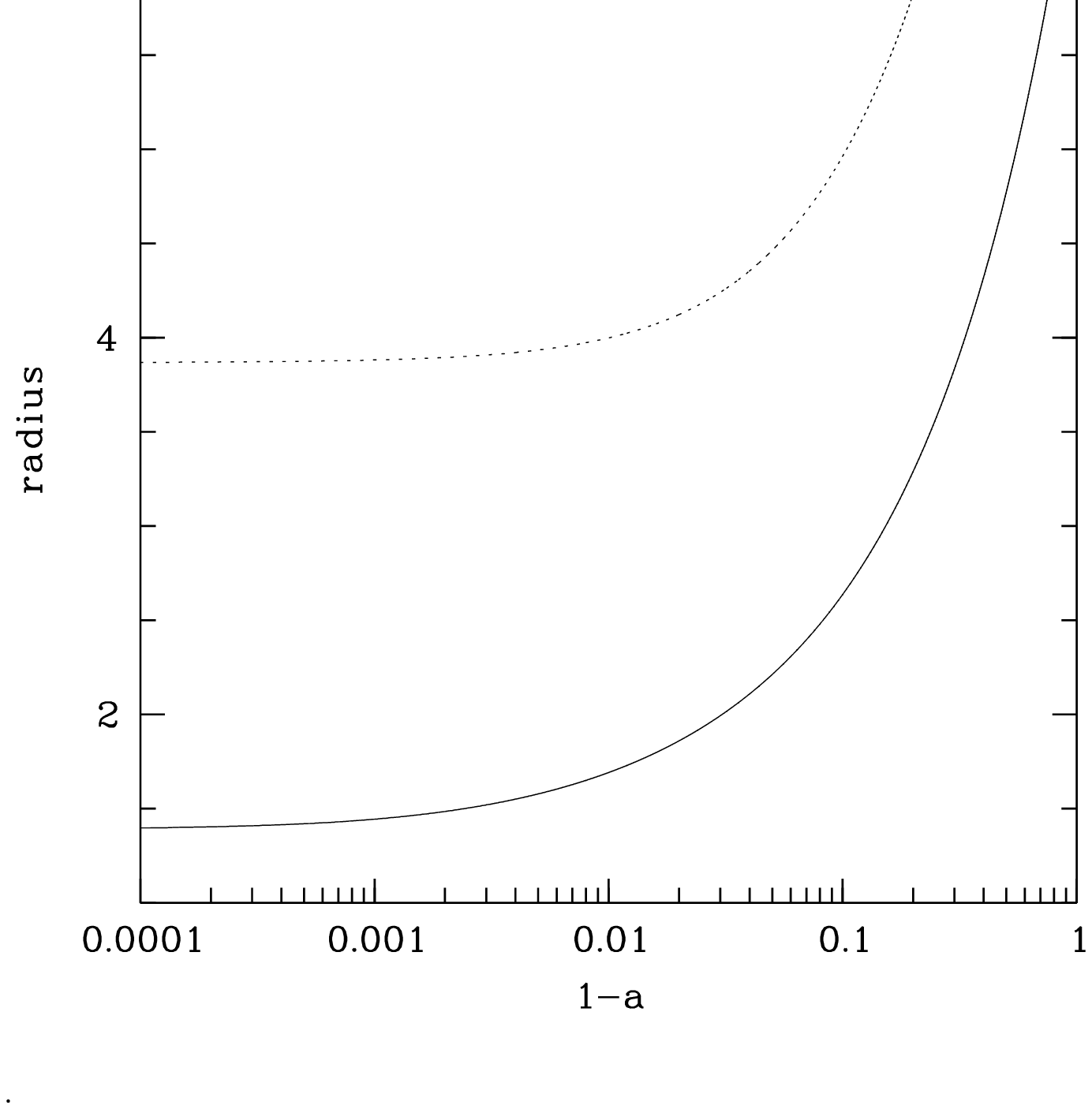,width=8cm,angle=0,%
bbllx=75pt,bblly=420pt,bburx=470pt,bbury=815pt,clip=}
\caption{Radii of orbits for which $\Omega\sb{V}$/$\Omega\sb{R}$ = 3:1 (solid line) 
or $\Omega\sb{V}$/$\Omega\sb{R}$ = 3:2 (dashed line) 
versus $(1-a)$. 
\vspace*{-0.4cm}
}
\label{fig1}
\end{figure}

\begin{figure}[t]
\psfig{figure=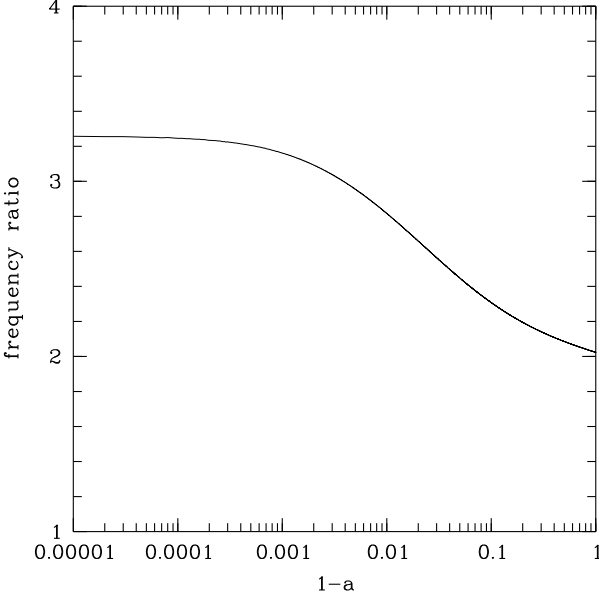,width=8cm,angle=0,%
bbllx=75pt,bblly=420pt,bburx=470pt,bbury=815pt,clip=}
\caption{Ratio of Kepler frequencies $\Omega\sb{\rm K}(r=r\sb{31}, a)$/$\Omega\sb{\rm K}(r=r\sb{32},a)$ versus $(1-a)$.
\vspace*{-0.4cm}
}
\label{fig2}
\end{figure}

The model I propose to explain the twin HFQPOs observed at a ratio of 3:2 or 3:1 should have both 
$\Omega\sb{\rm V}$/$\Omega\sb{\rm R}$ = 3:2 and $\Omega\sb{\rm V}$/$\Omega\sb{\rm R}$ = 3:1 for the same angular momentum $a$. Of course, 
this requires two different orbits with different radii 
called $r\sb{32}$ and $r\sb{31}$, which follow from Equations \ref{eq:1} to \ref{eq:3}. 
Figure~\ref{fig1} shows $r\sb{32}$ and $r\sb{31}$ versus $(1-a)$. There is no limitation on $a$, and $r\sb{32}$ and $r\sb{31}$ exist 
for any $a$. The second requirement is that the two orbits should be commensurable orbits, which in traditional sense means 
that $\Omega\sb{\rm K}(r=r\sb{31})$ = $n\times\Omega\sb{\rm K}(r=r\sb{32})$ in it's simplest form, with $n$ a natural number. 
Figure~\ref{fig2} shows $\Omega\sb{\rm K}(r=r\sb{31},a)$/$\Omega\sb{\rm K}(r=r\sb{32},a)$ versus $(1-a)$. The maximum value 
of this ratio is 3.26 and the minimum value is 2.026. Therefore there is a solution but only one solution for $n=3$, which in 
turn fixes the values of $a$ to $a=0.99616$, $r\sb{31}=1.546$ and $r\sb{32}=3.919$. Values of $a$,  $r\sb{31}$ and 
 $r\sb{32}$ are truncated.  

Table~\ref{fresum} summarizes the values of the relevant frequencies $\Omega$ including the Lense-Thirring precession frequency 
$\Omega\sb{\rm{LT}} = \Omega\sb{\rm K} - \Omega\sb{\rm V}$
(Merloni et al., \cite{MVSB1999}) for r$\sb{31}$ and r$\sb{32}$. The parametric resonances are for ratios of 3:1 and 3:2. I note 
that if a beat frequency of $\Omega\sb{\rm b} = \Omega\sb{\rm V} - \Omega\sb{\rm R}$ exists, which 
may be indicated by the 
observation of the 92 Hz frequency in XTE J1550-564, then the parametric resonances would be a 3:2:1 triplet both at 
$r\sb{31}$ and $r\sb{32}$. Therefore I include $\Omega\sb{\rm b}$ in Table~\ref{fresum}. 

\begin{table}
   \caption[]{Compilation of frequency values for $a=0.99616$, $r\sb{31}=1.546$ and
              $r\sb{32}=3.919$.}
    \label{fresum}
\begin{flushleft}
\begin{tabular}{ccc}
            \noalign{\smallskip}
\hline
\hline
$\Omega$&$r\sb{31}$&$r\sb{32}$ \\
\hline
\noalign {\smallskip}
$\Omega\sb{\rm K}$&0.342684&0.114228 \\
\noalign {\smallskip}
$\Omega\sb{\rm V}$&0.142370 &0.0942112\\
\noalign {\smallskip}
$\Omega\sb{\rm R}$&0.0475116&0.0628118\\
\noalign {\smallskip}
$\Omega\sb{\rm b}$&0.0948583&0.0313993 \\
\noalign {\smallskip}
$\Omega\sb{\rm{LT}}$&0.200313&0.0200171\\
\hline
\end{tabular}
\end{flushleft}  
\end{table}

It is evident from Table ~\ref{fresum} that $\Omega\sb{\rm V}$ is the upper frequency both in the 3:2 and the 3:1 resonance between 
$\Omega\sb{\rm V}$ and $\Omega\sb{\rm R}$ neglecting  $\Omega\sb{\rm K}$. 
Accordingly, the black hole mass can be calculated  for the 
3:1 resonance ($M\sb{31}$) and the 3:2 resonance ($M\sb{32}$) from Equations \ref{eq:4} and \ref{eq:5} based on the HFQPO measured upper frequency 
$\nu\sb{up}$ in Hz:

\begin{equation}\label{eq:4}
$$M\sb{31}/M\sb{\odot} = 4603.3/\nu\sb{up}$$
\end{equation}  

\begin{equation}\label{eq:5}
$$M\sb{32}/M\sb{\odot} = 3046.2/\nu\sb{up}$$
\end{equation}

Equations \ref{eq:4} and \ref{eq:5} illustrate that $M\sb{31}$ and $M\sb{32}$ differ by factor of 
1.51 only. 

Apart from the parametric resonances enforced by the initial requirements, i.e.  
$\Omega\sb{\rm V}(r=r\sb{31})$/$\Omega\sb{\rm R}(r=r\sb{31})$ = 3:1,
$\Omega\sb{\rm V}(r=r\sb{32})$/$\Omega\sb{\rm R}(r=r\sb{32})$ = 3:2 
and 
$\Omega\sb{\rm K}(r=r\sb{31})$/$\Omega\sb{\rm K}(r=r\sb{32})$ = 3:1 
it turns out that there are 'cross' resonances connecting $r\sb{31}$ and $r\sb{32}$, which are, 
apart from  the Kepler commensurable orbits,  
$\Omega\sb{\rm R}(r=r\sb{32})$/$\Omega\sb{\rm R}(r=r\sb{31})$ = 4:3 and  
$\Omega\sb{\rm V}(r=r\sb{32})$/$\Omega\sb{\rm R}(r=r\sb{31})$ = 2:1 (c.f. Table~\ref{fresum}). 
In summary, $r{\sb{31}}$ and $r{\sb{32}}$ show an extraordinary high degree of parametric resonances.  
   
With $a=0.99616$ some radii, useful to know, can be calculated. The event horizon is at 
$r\sb{H}=1.088$; the static limit in the equatorial plane is at $r\sb{0}=2$ and 
the marginally stable orbit is at $r\sb{ms}=1.25$ (defined by $\Omega\sb{R} =0$; for the definition by Bardeen et al. (1972)
$r\sb{ms}=1.306$),
 which in either case means that both $r\sb{31}$ and $r\sb{32}$ are stable orbits, and that 
$r\sb{31}$ is within the ergosphere whereas $r\sb{32}$ is outside of it. 
In fact, $r\sb{31}=(r\sb{H}+r\sb{0})/2$, which is mid-way between the event horizon and the static limit in the 
equatorial plane.

\section{Comparison with observations}

The model so far has made use of just frequency ratios not so much of the absolute frequencies. The test of the model 
would consist of a comparison with otherwise measured masses or angular momenta. For the three microquasars 
GRO J1655-40, XTE J1550-564 and GRS 1915+105 measurements of their masses have been made but there is so far no way 
to determine the angular momenta, so the test is necessarily restricted to their masses. Masses are usually derived 
from the determination of the orbit parameters in case of a binary. This holds for the three microquasars but also 
for the black hole Sgr\,A* in the Galactic Center which is orbited by the  S2 (S0-2) star
(Sch\"odel et al., 2002, Ghez et al., 2003). Genzel et al. (2003) have reported a QPO period 
of 16.8\, min observed in two near-infrared flares. Aschenbach et al. (2004) have claimed additionally 
quasi-periods 
around 219\, s, 700\, s, 1150\, s and 2250\, s, with the 1150\, s period being consistent with the 
NIR period. 
This set of quasi-periods was found in the power density spectra of one X-ray flare observed with {\sl{Chandra}} 
(Baganoff et al., 2001) and a second X-ray flare with {\sl{XMM-Newton}} (Porquet et al., 2003). 
Interesting in this context is that the frequencies corresponding to the latter three 
quasi-periods are close to a 3:2:1 ratio. Enforcing such a ratio sequence in a best fit, it turns out that such a ratio 
is consistent with the measurements. Accordingly, I include the Galactic Center black hole in this test 
and predict the mass of Sgr\,A*.

\subsection{\it{GRO J1655-40}}

The frequencies in question are 450 and 300 Hz (Strohmayer 2001a, Remillard et al. 1999) with 
$\Omega\sb{V}$ = 450 Hz. According to Equation \ref{eq:5} the predicted black hole mass is 
$M\sb{BH} = 6.76 \pm 0.1 M\sb{\odot}$. The relative uncertainty of the mass is the same as that  
of the HFQPO measurements of $\approx$ 1.5\%. Dynamical mass measurements have been reported 
by Orosz \& Bailyn (1997) with 
$M\sb{BH} = 7.02\pm 0.22 M\sb{\odot}$ and more recently by Greene et al. (2001) who obtained 
a mass range of $M\sb{BH} = 5.8 - 6.8 M\sb{\odot}$. The agreement between model prediction 
and observation is quite satisfactory.

\subsection{\it{XTE J1550-564}}

Frequencies of 276 and 184 Hz (Remillard et al., 2002) have been measured. With 
$\Omega\sb{V}$ = 276 Hz Equation \ref{eq:5} predicts a black hole mass of 
$M\sb{BH} = 11.04 \pm 0.2 M\sb{\odot}$. The relative uncertainty of the mass is again due to the accuracy
of the HFQPO measurements of $\approx$ 2\%. Dynamical mass measurements have been reported
by Orosz et al. (2002a) with a $\pm$1$\sigma$ mass range of 
$M\sb{BH} = 8.4 - 11.6 M\sb{\odot}$, which is nicely matched by the predicted mass.

\subsection{\it{GRS 1915+105}}

The relevant 3:2 pair at 168 and 113 Hz has only recently been found (Remillard et al., 2003, McClintock \& Remillard, 2004). 
With $\Omega\sb{V}$ = 168 Hz Equation \ref{eq:5} predicts a black hole mass of
$M\sb{BH} = 18.13 \pm 0.36 M\sb{\odot}$. The relative uncertainty of the mass is due the accuracy
of the HFQPO measurements of $\approx$ 2\%. Dynamical mass measurements have been reported
by Greiner at al. (2001) with $M\sb{BH} = 14.0 \pm 4.0 M\sb{\odot}$ and Harlaftis \& Greiner (2004) with 
$M\sb{BH} = 14.0 \pm 4.4 M\sb{\odot}$. 
Also in this case the predicted mass matches the dynamically mass within the $\pm$1$\sigma$ range, but the measurements of this source 
illustrate nicely the potential uncertainty in dynamical mass measurements. Harlaftis \& Greiner (2004) point out that the biggest uncertainty 
is related to the inclination of the orbit, which for GRS 1915+105 is taken from the inclination of the associated 
jet assuming that the jet axis is orthogonal to the orbital plane of the binary. If, for instance, the inclination angle is changed from the 
adopted 66$\sp{\deg}$ to 56$\sp{\deg}$ the estimated black hole mass is 
$M\sb{BH} = 16.9 \pm 5.9 M\sb{\odot}$ (Harlaftis \& Greiner, 2004). Actually, Kaiser et al. (2004) suggest that the inclination angle is 
53$\sp{\deg}$ putting the best estimate for the dynamical black hole mass at 17.8 $M\sb{\odot}$, which is within the $\pm$1$\sigma$ measurement 
of  Harlaftis \& Greiner (2004), but it is also pretty close to the predicted value of $18.1 M\sb{\odot}$.

\subsection{\it{XTE J1859-226}}

Cui et al. (2000) report the measurement of a single HFQPO at 187 Hz associated with XTE J1859-226. 
Filippenko \& Chornock (2001) announced a mass function of the binary of $7.4 M\sb{\odot}$, and 
Brocksopp et al. (2002) discovered associated radio jets suggesting  that XTE J1859-226 is another galactic microquasar. 
Orosz et al. (2002b) and McClintock \& Remillard (2004) finally quote an estimated mass of the black hole  
between 7.6 and 12.0 $M\sb{\odot}$. So far there is just one HFQPO measured so that a prediction based on the 
proposed model is difficult, but there are only two options. If the 187 Hz frequency corresponds to the vertical epicyclic 
3:2 resonance the predicted mass (Equation \ref{eq:5}) is about 16.3 $M\sb{\odot}$ and a second HFQPO at 125 Hz, i.e. 
the radial epicyclic frequency, should be looked for. If, on the contrary, the 187 Hz frequency is the result of the 
radial epicyclic frequency the expected vertical epicyclic frequency is at 281 Hz and the predicted 
mass is about 10.9  $M\sb{\odot}$, which fits quite nicely in the mass range given by Orosz et al. (2002b) and 
McClintock \& Remillard (2004), but these authors mark the mass range as quite uncertain. 
A deep search for a second HFQPO would be extremely useful.

\subsection{The Galactic Center black hole Sgr\,A*}

Aschenbach et al. (2004) have grouped the frequencies indicated in the power density spectra 
of two  
NIR flares (Genzel et al. 2003) and two X-ray flares. There are 
three groups for which there is evidence that each of the frequencies of one group 
occurred in at least one NIR and both X-ray flares. 
These three  groups  of quasi-periods in seconds, ordered by NIR, {\sl{Chandra}}, 
{\sl{XMM-Newton}},  are  group 1: (733, 692, 701), 
group 2: (1026, 1117, 1173) and group 3: (-, 2307, 2178); the duration of the NIR flare was too 
short to cover the longest period. 
Looking at the average quasi-period  within one group it appears as if the ratio of the average quasi-periods follows 
a 1:1.5:3 relation which is equivalent to a 3:2:1 relation in frequencies. This is at least consistent 
in view of the frequency measurement uncertainty.     
Each of these eight quasi-periods has been observed at low wave number which means that the 
systematic uncertainty of each period determined from the associated power density spectra is up to $\pm$20\%.  
Enforcing a 3:2:1 ratio sequence in a best fit the highest frequency is $\nu\sb{up}$ = 1.4035 $\pm$ 0.0574 mHz or a period 
of 712$\pm$29 s. The two remaining quasi-periods are at 1069$\pm$44 s and 2138$\pm$87 s. 
In the paper of Aschenbach et al. (2004) we associated the quasi-period of group 1 with the 
vertical epicyclic oscillation at the marginally stable orbit and the quasi-period of group 2 with the 
radial epicyclic oscillation at the radius at which the radial oscillations have their maximum frequency. 
This together with the association of the 219 s quasi-period to the Kepler period of the marginally stable 
orbit led to a unique value for the mass and the angular momentum of the black hole. For the quasi-period 
of group 3 (2150 s) we had no convincing suggestion. This together with the fact that we needed to employ different 
radial positions for the different frequencies is unsatisfactory. This is taken care now by the association 
of the quasi-periods of group 1, group 2 and group 3 with the epicyclic modes at just one radius.

Unlike the microquasars which show two frequencies Sgr\,A* appears to show three frequencies of about similar power 
spectral densities (Aschenbach et al., 2004), so that it makes sense to classify this as an example of the 3:1 resonance 
for which Equation \ref{eq:4} applies, and the black hole mass of Sgr\,A* is predicted to 
$M\sb{BH} = (3.28 \pm 0.13)\times$10$\sp 6 M\sb{\odot}$. 
Measurements of the orbital motion of the star S0-2 (S2) around the GC black hole have
 resulted in M$\sb{\rm{BH}}$ = 3.7$\pm$1.5 $\times$ 10$\sp 6$M$_\odot$ (Sch\"odel et al., \cite{S2002}), 
M$\sb{\rm{BH}}$ = 4.07$\pm$0.62 $\times$ 10$\sp 6$M$_\odot$~(R$\sb 0/8 {\rm{kpc}})\sp 3$
 (Ghez et al., \cite{Gh2003}) with R$\sb 0$ the distance to the black hole, and 
M$\sb{\rm{BH}}$ = 3.59$\pm$0.59 $\times$ 10$\sp 6$M$_\odot$ (Eisenhauer et al., \cite{ESG2003}). 
Errors quoted for each of the three mass measurements are 1$\sigma$. 
The mass predicted by the present model 
agrees with both the Sch\"odel et al. (2002) mass and the mass given by 
Eisenhauer et al. (2003) but it is slightly out of the 1$\sigma$ range of the Ghez et al. 
(2003) measurement 
for $R = 8 \rm{kpc}$.  
 
Application of Equation \ref{eq:5}, checking for a 3:2 resonance, gives 
$M\sb{BH} = (2.17 \pm 0.1)\times$10$\sp 6 M\sb{\odot}$, which is ruled out by the dynamical measurements. 
This supports that in the framework of this model  
the accretion disk around Sgr\,A* is in a 3:1 resonance. 
The third frequency associated with the resonance state 2 may be either a beat frequency, i.e. $\Omega\sb{\rm{V}}$-$\Omega\sb{\rm{R}}$ or the 
first harmonic of $\Omega\sb{\rm{R}}$. In either case a fourth frequency could be expected with is either $\Omega\sb{\rm{V}}$+$\Omega\sb{\rm{R}}$ 
or the third harmonic of $\Omega\sb{\rm{R}}$. This would correspond to resonance state 4, and the expected period is  
534$\pm$22 s. Such a period seems to be present in the power density spectrum of the {\sl{Chandra}} flare at 494$\pm$41 s and in the NIR 
 flare of June 15, 2003 at 
498$\pm$100 s. It is absent in the NIR flare of June 16, 2003 and the {\sl{XMM-Newton}} flare (Aschenbach et al., 2004). 
With $a$ and $M\sb{BH}$ given the orbital (Kepler) period is P$\sb{K}(r=r\sb{31})=296\pm12$ s; at the marginally stable orbit 
P$\sb{K}(r=r\sb{ms})=242\pm10$ s. The {\sl{Chandra}} flare shows increased power spectral density at P$\sb{Chandra} = 256\pm 10$ s and the 
{\sl{XMM-Newton}} flare has significant power at P$\sb{XMM} = 219\pm 15$ s (Aschenbach et al., 2004). These data are consistent with the view 
that the shortest quasi-periods observed in the X-ray flares correspond to Keplerian motion along orbits between $r\sb{31}$ and $r\sb{ms}$, the actual 
radius might be different from flare to flare, not so are $\Omega\sb{V}$ and  $\Omega\sb{R}$, of course.   

\section{A physical mechanism?}
 
With the distinct values for $a$, $r\sb{31}$ and $r\sb{32}$ given it comes to mind to look for a physical reason why these quantities are special. 
I checked the general relativistic expressions of energy, angular momentum and orbital frequency for a test particle moving around a  
rotating black hole 
on  a stable orbit. These expressions (Shapiro \& Teukolsky, 1983) change monotonically with $r$ 
in ($a$, $r$) space. But this is not the case 
for the orbital velocity $v\sp{(\Phi)}$ described in the ZAMO-frame (the {\it{Zero Angular Momentum Observer}} or {\it{Bardeen observer}}). 
$v\sp{(\Phi)}$ is represented by Equation \ref{eq:6} (M\"uller \& Camenzind, 2004): 

\begin{equation}\label{eq:6}
$$v\sp{(\Phi)} = {{{\tilde\omega}}\over{\alpha}}(\Omega -\omega)$$.
\end{equation}

The Boyer-Lindquist functions are (e.g., M\"uller \& Camenzind, 2004):

\begin{equation}\label{eq:7}
$$\alpha = {{\rho\sqrt{\Delta}}\over{\Sigma}}$$,
\end{equation}

\begin{equation}\label{eq:8}
$$\Delta = r\sp{2} -2Mr+a\sp{2}$$,
\end{equation}

\begin{equation}\label{eq:9}
$$\rho\sp{2} = r\sp{2}+a\sp{2} cos\sp{2} \theta$$,
\end{equation}

\begin{equation}\label{eq:10}
$$\Sigma\sp{2} = (r\sp{2}+a\sp{2})\sp{2}-a\sp{2}\Delta sin\sp{2}\theta$$,
\end{equation}

\begin{equation}\label{eq:11}
$${\tilde\omega} = {{\Sigma}\over{\rho}} sin \theta$$,
\end{equation}

\begin{equation}\label{eq:12}
$$\omega = {{2aMr}\over{\Sigma\sp{2}}}$$,
\end{equation}

$\theta$ is the polar angle with $\theta = \pi/2$ defining the equatorial plane. 

\begin{figure}[t]
\psfig{figure=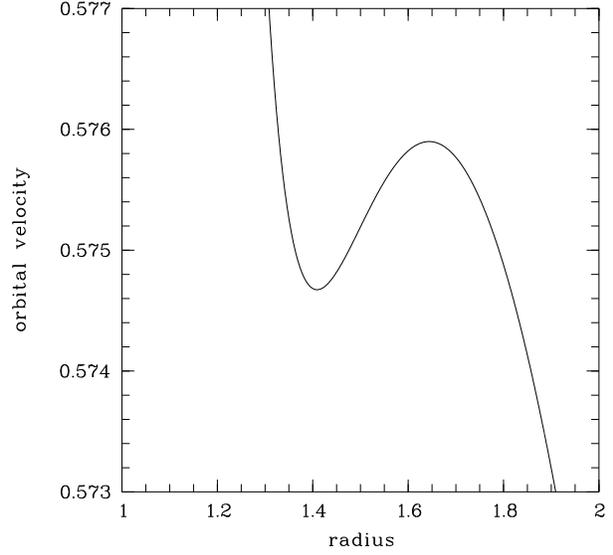,width=8cm,angle=0,%
bbllx=75pt,bblly=420pt,bburx=470pt,bbury=815pt,clip=}
\caption{Orbital velocity $v\sp{(\Phi)}$ versus orbital radius for 
$a=0.99616$. Outside the radial range shown $v\sp{(\Phi)}$ decreases monotonically 
with radius.
\vspace*{-0.4cm}
}
\label{fig3}
\end{figure}

\begin{figure}[t]
\psfig{figure=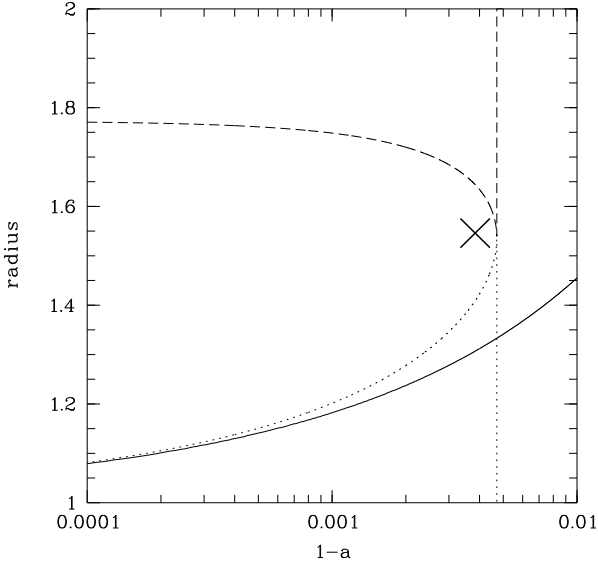,width=8cm,angle=0,%
bbllx=75pt,bblly=420pt,bburx=470pt,bbury=815pt,clip=}
\caption{As a function of $(1-a)$ are shown the radial position $r\sb{max}$ of the local maximum of $v\sp{(\Phi)}$ 
(curved dashed line),  
the radial position $r\sb{min}$ of the local minimum of $v\sp{(\Phi)}$ 
(curved dotted line). The vertical dashed line is 
at $a=a\sb{c}=0.9953$. The cross marks $r\sb{31}$ and $a=a\sb{f}=0.99616$, which is the solution of the parametric resonance model 
with commensurable orbits. The solid line represents the innermost marginally stable orbit $r\sb{ms}$ defined by $\Omega\sb{R}=0$.
\vspace*{-0.4cm}
}
\label{fig4}
\end{figure}

As M\"uller and Camenzind (2004) state one usually assumes Keplerian rotation 
for $r\ge r\sb{ms}$ and sets $\Omega = \Omega\sb{K}$ (Equation \ref{eq:1}). 
Setting $\theta = \pi/2$ and $\Omega = \Omega\sb{K}$ I have solved numerically Equation \ref{eq:6} over the  full ($a$, $r$) space. 
For large values of $r$ and small values of $a$, $v\sp{(\Phi)}$ behaves as expected from Newtonian mechanics, i.e. $v\sp{(\Phi)}$ increases 
monotonically with decreasing $r$. But this monotonic behaviour changes for ${a~>~0.9953}$; $v\sp{(\Phi)}$ develops a maximum at small 
values of $r$, decreases with decreasing $r$ until it reaches a minimum and from thereon increases again with decreasing $r$ until 
$r\sb{ms}$ is reached. Figure~\ref{fig3} shows an example for ${a~=~0.99616}$. The depression of $v\sp{(\Phi)}$ of 0.2\% is small but it increases 
with increasing $a$.  Figure~\ref{fig4} illustrates how the radial position of the local maximum ($r\sb{max}$) and 
the local minimum ($r\sb{min}$) of $v\sp{(\Phi)}$ change with $a$. The plot also demonstrates that always $r\sb{min}>r\sb{ms}$. 
Such a variation with $r$ is absent for the particle energy E and its angular momentum l. 
A main role, that this is different for the orbital velocity, 
seems to play the appearance of $\omega$, the frame-dragging frequency or potential for angular momentum, in Equation~\ref{eq:6}.  
The change of $v\sp{(\Phi)}$ implies that for $r\sb{min}\le r\le r\sb{max}$ 
$\partial E/\partial v\sp{(\Phi)}\le 0$ 
and $\partial l/\partial v\sp{(\Phi)}\ge 0$.  

\begin{figure}[t]
\psfig{figure=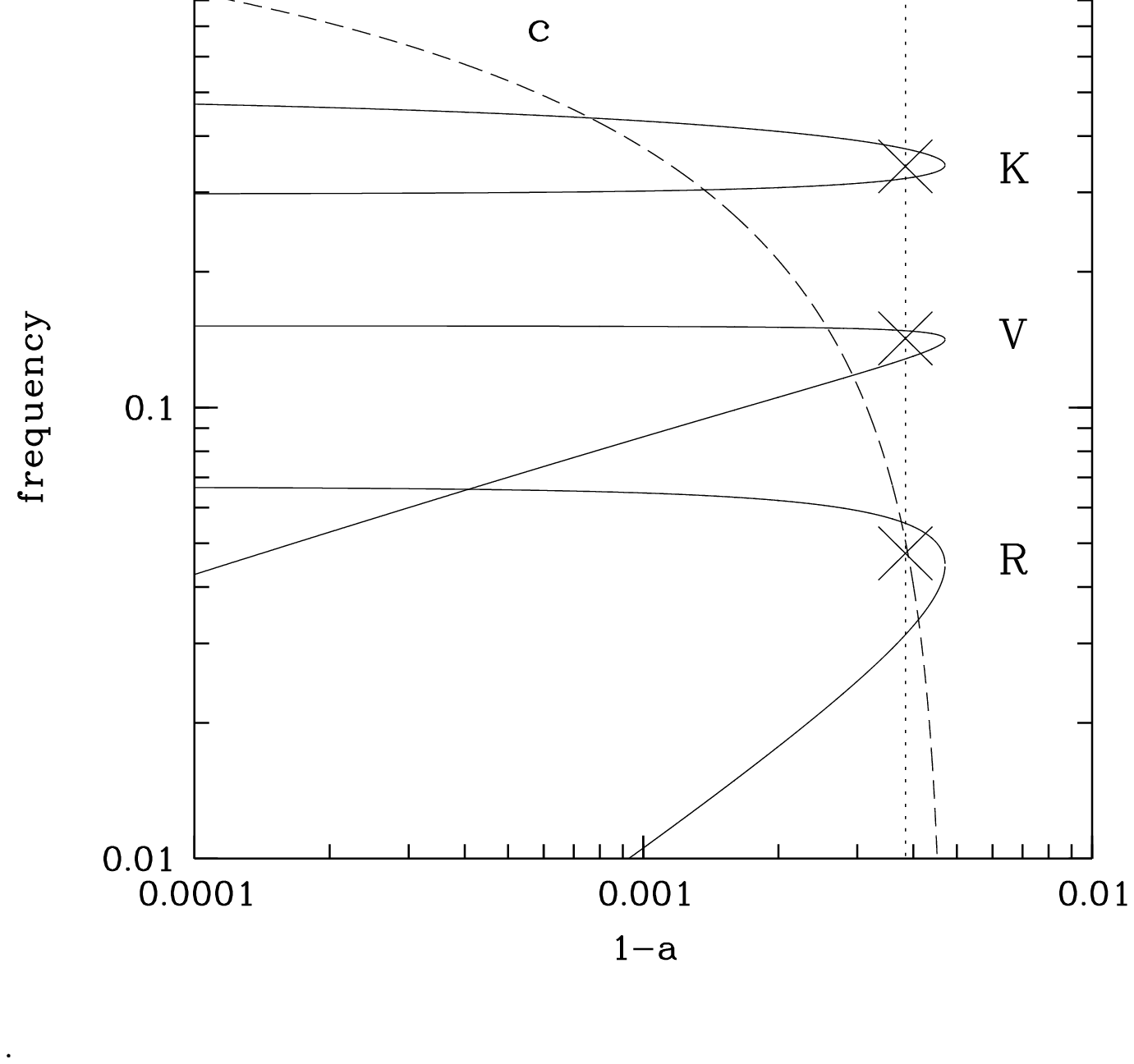,width=8cm,angle=0,%
bbllx=75pt,bblly=420pt,bburx=470pt,bbury=815pt,clip=}
\caption{
As a function of $(1-a)$ are shown $\Omega\sb{K}(r=r\sb{min}, a)$ and $\Omega\sb{K}(r=r\sb{max}, a)$ 
(label K), which makes up the upper parabola like contour; the contour labelled V is for 
$\Omega\sb{V}(r=r\sb{min}, a)$ and $\Omega\sb{V}(r=r\sb{max}, a)$ and the lowest parabola like contour (R) is 
for $\Omega\sb{R}(r=r\sb{min}, a)$ and $\Omega\sb{R}(r=r\sb{max}, a)$. The vertical dashed line is for $a=a\sb{f}=0.99616$ and the crosses mark 
the corresponding frequencies found from the commensurability considerations. Their positions are clearly within the radial section 
bounded by $r\sb{min}$ and $r\sb{max}$. The dashed line labelled c represents 
$\Omega\sb{c} = 2\pi {{\partial{v\sp{(\Phi)}}}\over{\partial r}}\mid\sb{max}$. 
\vspace*{-0.4cm}
}
\label{fig5}
\end{figure}

According to standard 
notion the orbits between $r\sb{min}$ and $r\sb{max}$ are stable. The region $r\sb{min}<r<r\sb{max}$ is the only region 
with $\partial{v\sp{(\Phi)}}/\partial r \ge 0$; it also has $\partial\sp{2}{v\sp{(\Phi)}}/\partial r\sp{2}=0$. 
It is basically the rate of change of the orbital velocity per 
unit length of $r$ seen by the {\it{ZAMO}}. It has  
the dimension of a frequency. I therefore define an angular 'frequency' $\Omega\sb{c} = 2\pi {{\partial{v\sp{(\Phi)}}}\over{\partial r}}\mid\sb{max}$, 
by which the radial position is defined at which $\partial{v\sp{(\Phi)}}/\partial r$ has the maximum value for a given $a$.     
Figure~\ref{fig5} shows $\Omega\sb{c}$ in comparison with $\Omega\sb{R}$, $\Omega\sb{V}$ and $\Omega\sb{K}$ versus (1-$a$). 
It might be a chance coincidence but $\Omega\sb{c}$ has the same value as $\Omega\sb{R}$ for ${a~=~0.99616}$ at  
$r\sb{31}$, indicated by the lowest cross in Figure~\ref{fig5}. It appears that there is a 1:1 resonance between 
$\Omega\sb{c}$ and $\Omega\sb{R}$, and this is the only 'resonance' with $\Omega\sb{R}$, $\Omega\sb{V}$ and $\Omega\sb{K}$  
for $a<0.998$, which is the 'Thorne' limit (Thorne, 1974). In summary, $\Omega\sb{c} = 2\pi {{\partial{v\sp{(\Phi)}}}\over{\partial r}}\mid\sb{max}$ 
picks the same values for $a$ and $r$ as the parametric resonance model with commensurable Kepler orbits.

\section{Discussion}

The model predicts the mass of each of the four black hole objects with 
well estimated masses quite satisfactorily from measured HFQPOs. Nevertheless, there are a 
couple of interesting coincidencies which I would like to point out and discuss. 

It is interesting to note that the model based on parametric resonances between the vertical and radial epicyclic frequencies constrained by the requirement 
that any associated orbits should be commensurable orbits has only one solution, which selects just two orbits with 
radii $r\sb{31}$ and $r\sb{32}$, and restricts the angular momentum of the black hole to just one value of $a=a\sb{f}=0.99616$. 
Even more interesting is the result that the orbital velocity $v\sp{(\Phi)}$ for the {\it{ZAMO}} changes its otherwise monotonical behaviour with 
radius and shows   
a local minimum and a local maximum which define a narrow radial annulus with $\partial{v\sp{(\Phi)}}/\partial r \ge 0$ for 
${a~>~0.9953=a\sb{c}}$. To me it seems that this feature of very rapidly rotating black holes has been overlooked so far. 
This certainly needs more study.  
Furthermore, the fact that $\Omega\sb{c} = 2\pi {{\partial{v\sp{(\Phi)}}}\over{\partial r}}\mid\sb{max} = 
\Omega\sb{R}(r=r\sb{31})$, suggests that there is a link between the 'strange' behaviour of $v\sp{(\Phi)}$ and the radial epicyclic oscillation, such 
that $\Omega\sb{c}(a\sb{f}) =
\Omega\sb{R}(r\sb{31}, a\sb{f})$ is the condition to excite the radial and vertical epicyclic oscillations. I note that so far all this is produced in 
a single test particle environment. 

The model predicts that there is just one single value of $a=a\sb{f}$, and the HFQPOs of the three microquasars and the GC black hole are consistent 
with this. The same value of $a$ for four sources suggests that this is some sort of final state of $a$; if correct this would 
replace the former 'Thorne' upper limit of $a=0.9982-0.9978$ (Thorne, 1974). This would also imply that the excitation of the radial and vertical  
epicyclic oscillations at $r\sb{31}$  somehow regulate the angular momentum flow towards the black hole to an equilibrium point.   
$r{\sb{31}}$ is located in the ergosphere halfway between the horizon and the static limit; the distance $\delta r$ to the ergosphere limit is 
$\delta r=0.45$ in the equatorial plane. A radial oscillation with an amplitude greater than $\delta r$ would traverse the ergosphere boundary 
back and forth.

The oscillations at $r\sb{31}$ are likely to be powered by mass accretion and one could envisage the following scenario. At low 
mass accretion rates both the vertical and radial oscillations are stable and energy is stored in the corresponding wave; with increasing 
accretion rate the vertical oscillations become unstable because they are not any longer capable to store the energy and/or 
angular momentum but the 
energy is released in flares or bursts. I suggest that this is the state in which the GC black hole is. 
Observations show that the
quiescent state X-ray luminosity of Sgr\,A* is about 11 orders of magnitude below the Eddington limit
(Baganoff et al., 2001, Porquet et al., 2003), which
might indicate that the accretion rate is fairly low, although other mechanisms for the low luminosity have been proposed.
The time scale for the rise time of the flare should be a half up to a few times of the period of the vertical oscillations, i.e. about 350 to 1000 s, 
 and the 
duration of such a flare should be about a few times of the period of the radial oscillations, i.e. 40 to 120 min. These timescales 
have been observed in the {\sl{Chandra}} (Baganoff et al., 2001) flare and the {\sl{XMM-Newton}} flare (Porquet et al., 2003, Aschenbach et al., 2004). 
Recently Bower et al. (2004) have reported
measurements by which Sgr\,A* has been spatially resolved at a wavelength of 7 mm, and they estimate itssize to be 
$<$2.4$\times$10$\sp{12}$ cm at a wavelength of 1.3 mm for a distance of 8 kpc. 
I note that the diameter of the region, where the vertical epicyclic
mode is at work, is 2$\times r\sb{31}$ = 1.5$\times$10$\sp{12}$ cm, whereas the diameter of the $r\sb{32}$ annulus is 
 3.8$\times$10$\sp{12}$ cm, which appears to be incompatible with the wavelength extrapolated radio measurements, if the X-ray
emitting region is not more extended than the radio source. This result seems to confirm that the Sgr\,A* emission region
is restricted to an area with a radius $r\le r\sb{31}$.

    At even greater accretion rates 
there are no stable vertical oscillations any more, the 3:1 resonance disappears, and probably a significant 
fraction of the energy of the accreting matter is lost in vertical direction in a non-periodic fashion or 
stored in the radial mode, the amplitude of which will grow. Eventually the radial wave will reach $r\sb{32}$ and power 
the vertical and radial modes at $r\sb{32}$ because of their parametric 'cross' resonances (section 2). This is the 3:2 state 
I suggest the three microquasars are in. The X-ray luminosities of the three microquasars are very high, close to or possibly slightly above 
the Eddington limit. 
The accretion rate is certainly fairly high. Eventually, outbursts may also occur in the 3:2 state, when the 
vertical mode oscillations become unstable temporarily. Of course, a quantitative  relation between 
oscillation state and accretion rate is still to be worked out. 
 
The three microquasars are known to have jets, and Bower et al. (2004) state that their radio measurements 
of Sgr\,A* are consistent with jet models, so that 
all four sources might have jets. Therefore one might ask the question, whether the properties twin or triple HFQPOs, $a=a\sb{f}$ and 
existence of a jet come together all three, and vice versa. For instance, has  
an accreting black hole 
with a jet $a=a\sb{f}$ and HFQPOs? If so, it could be profitable to search for instance the extragalactic jet sources, make an estimate 
of the HFQPOs 
from Equation \ref{eq:4} and Equation \ref{eq:5}, and when found determine the black hole mass to percentage accuracy. 
Finally, though the idea might be foolish, it comes to mind. Is the instability of the vertical epicyclic oscillation 
at $r=r\sb{31}$ the 
seed for the creating of accretion driven black hole jets? The decrease of $v\sp{(\Phi)}$ of about 2\% between $r\sb{max}$ and 
$r\sb{min}$ would correspond to an apparent loss of orbital kinetic energy of about 4.3 MeV/nucleon in non-relativistic Newtonian 
approximation. Whether and how this energy can be made use of is unclear.

\section{Conclusion}

I have constructed a model built upon parametric 3:1 and 3:2 resonances between the vertical and radial epicyclic oscillations of 
a test particle orbiting a rotating black hole. If these two orbits are required to be commensurable orbits as far as 
their Keplerian frequencies are concerned there exists only one solution, which specifies uniquely the orbit radius 
$r\sb{31}=1.546$  
of 
the 3:1 resonance, the orbit radius 
$r\sb{32}=3.919$
of the 3:2 resonance and the angular momentum $a\sb{f}=0.99616$. 
The analysis of the general relativistic expression for the orbital velocity $v\sp{(\Phi)}$ in the {\sl{ZAMO}} frame shows 
a deviation from the monotonic relationship between $v\sp{(\Phi)}$ and $r$ for $a~>~0.9953=a\sb{c}$. 
$\partial{v\sp{(\Phi)}}/\partial r$ changes its sign over a narrow radial annulus, 
and there 
$\Omega\sb{c} = 2\pi {{\partial{v\sp{(\Phi)}}}\over{\partial r}}\mid\sb{max}=\Omega\sb{R}$. 
It seems that this condition is responsible for the excitation of radial and vertical epicyclic oscillations. 
The effect of $\partial{v\sp{(\Phi)}}/\partial r$ changing its sign is likely to be related to energy and angular momentum transfer via 
vertical and radial waves. This effect seems to have been overlooked in the past and needs more study. 
Since so far the model is based on test particle motion the entire effect is due to general relativity and for rapidly 
rotating black holes. It is not a collective effect.  
This, in some sense, consistent model predicts a new upper limit of $a=a\sb{f}=0.99616$, which is slightly lower than the 'Thorne' limit
(Thorne, 1974). The model predicts the mass of fully spun-up black holes on the basis of measured twin (3:2) or 
triple (3:1) HFQPOs to the accuracy of the frequency measurements. It is suggested that the 3:1 resonance is active at 
low accretion rates, and the 3:2 resonance can be observed at very high accretion rates. At intermediate 
accretion rates $\Omega\sb{R}(r=r\sb{31})$ may be the only persistent frequency. The comparison of the masses derived with this 
model, on the basis of the measured HFQPOs, with the masses derived from dynamical measurements, i.e. binary orbit 
characteristics, shows an excellent agreement for the three microquasars GRO J1655-40, XTE J1550-564 and GRS 1915+105 and lowers  
the mass uncertainty significantly. The frequencies measured recently by Aschenbach et al. (2004) for Sgr\,A* appear to 
be arranged in a 3:2:1 ratio and the derived mass, which is 5.7 orders of magnitude higher than the lowest mass of the 
microquasars involved, agrees quite well with the dynamical mass at a much reduced uncertainty.      

\begin{acknowledgements}
I like to thank Joachim Tr\"umper for inspiring discussions.
\end{acknowledgements}

           \end{document}